\documentclass[fleqn,usenatbib]{mnras}

\usepackage{newtxtext,newtxmath}

\usepackage[T1]{fontenc}
\usepackage[normalem]{ulem} 
\DeclareRobustCommand{\VAN}[3]{#2}
\let\VANthebibliography\thebibliography
\def\thebibliography{\DeclareRobustCommand{\VAN}[3]{##3}\VANthebibliography}

\usepackage{graphicx}	
\usepackage{amsmath}	
\usepackage{pdflscape}
\usepackage{subcaption}
\title[Properties of atomic hydrogen gas in the Galactic plane]{Properties of atomic hydrogen gas in the Galactic plane from THOR 21-cm absorption spectra: a comparison with the high latitude gas}

\author[Basu et al.]{
Arghyadeep Basu,$^{1,2}$\thanks{E-mail: basu.arghyadeep@yahoo.in}
Nirupam Roy,$^{3}$
Henrik Beuther,$^{4}$
Jonas Syed,$^{4}$
J\"urgen Ott, $^{5}$
\newauthor{
Juan D. Soler, $^{4}$
Jeroen Stil $^{6}$
and Michael R. Rugel $^{7}$ }
\\
\\
$^{1}$Max-Planck-Institut f$\ddot{u}$r Astrophysik, Karl-Schwarzschild-Straße 1, 85741 Garching, Germany\\
$^{2}$Department of Physics, Presidency University, Kolkata 700073,India\\
$^{3}$Department of Physics, Indian Institute of Science, Bangalore 560012, India\\
$^{4}$Max-Planck-Institut für Astronomie, K\"onigstuhl 17, 69117 Heidelberg, Germany\\
$^{5}$National Radio Astronomy Observatory, PO Box O, 1003 Lopezville Road, Socorro, NM 87801, USA\\
$^{6}$Department of Physics and Astronomy, The University of Calgary, 2500 University Drive NW, Calgary AB T2N 1N4, Canada\\
$^{7}$Max-Planck-Institut f$\ddot{u}$r Radioastronomie, Auf dem H$\ddot{u}$gel 69, 53121 Bonn, Germany\\
}

\date{Accepted 2022 October 17. Received 2022 September 25; in original form 2022 June 03}

\pubyear{2022}

\begin{document}
\label{firstpage}
\pagerange{\pageref{firstpage}--\pageref{lastpage}}
\maketitle

\begin{abstract}
The neutral hydrogen 21 cm line is an excellent tracer of the atomic interstellar medium in the cold and the warm phases. Combined 21 cm emission and absorption observations are very useful to study the properties of the gas over a wide range of density and temperature. In this work, we have used 21 cm absorption spectra from recent interferometric surveys, along with the corresponding emission spectra from earlier single dish surveys to study the properties of the atomic gas in the Milky Way. In particular, we focus on a comparison of properties between lines of sight through the gas disk in the Galactic plane and high Galactic latitude lines of sight through more diffuse gas. As expected, the analysis shows a lower average temperature for the gas in the Galactic plane compared to that along the high latitude lines of sight. The gas in the plane also has a higher molecular fraction, showing a sharp transition and flattening in the dust - gas correlation. On the other hand, the observed correlation between 21 cm brightness temperature and optical depth indicates some intrinsic difference in spin temperature distribution and a fraction of gas in the Galactic plane having intermediate optical depth (for $0.02<\tau<0.2$) but higher spin temperature, compared to that of the diffuse gas at high latitude with the same optical depth. This may be due to a small fraction of cold gas with slightly higher temperature and lower density present on the Galactic plane.

\end{abstract}




\section{Introduction}

The interstellar medium (ISM) consists mostly of hydrogen gas, in ionized, neutral atomic and molecular form. The 21 cm transition, originating from the hyperfine splitting due to spin-spin coupling in the neutral hydrogen atom (H~{\sc i}) in its ground state, is thus very useful in in studying the ISM of the Milky Way and of nearby galaxies, as well as probing the distant universe through the redshifted 21 cm line \citep[]{Clark,Field,Walter}. For the Galactic H~{\sc i}, the 21 cm signal can be detected either as emission line from the gas, or absorption line towards suitable background continuum sources. The line of sight column density ($N_{HI}$) and the number density distribution in the hyperfine (triplet and singlet) energy levels, specified by the spin temperature ($T_{s}$), determine the strength of emission and absorption. Furthermore, the linewidth of emission/absorption spectra is determined by the kinetic temperature ($T_{k}$) and non-thermal broadening. Thus, H~{\sc i} 21 cm observations can be used to determine various physical properties of the gas.

\begin{figure*}
	\includegraphics[width=100mm]{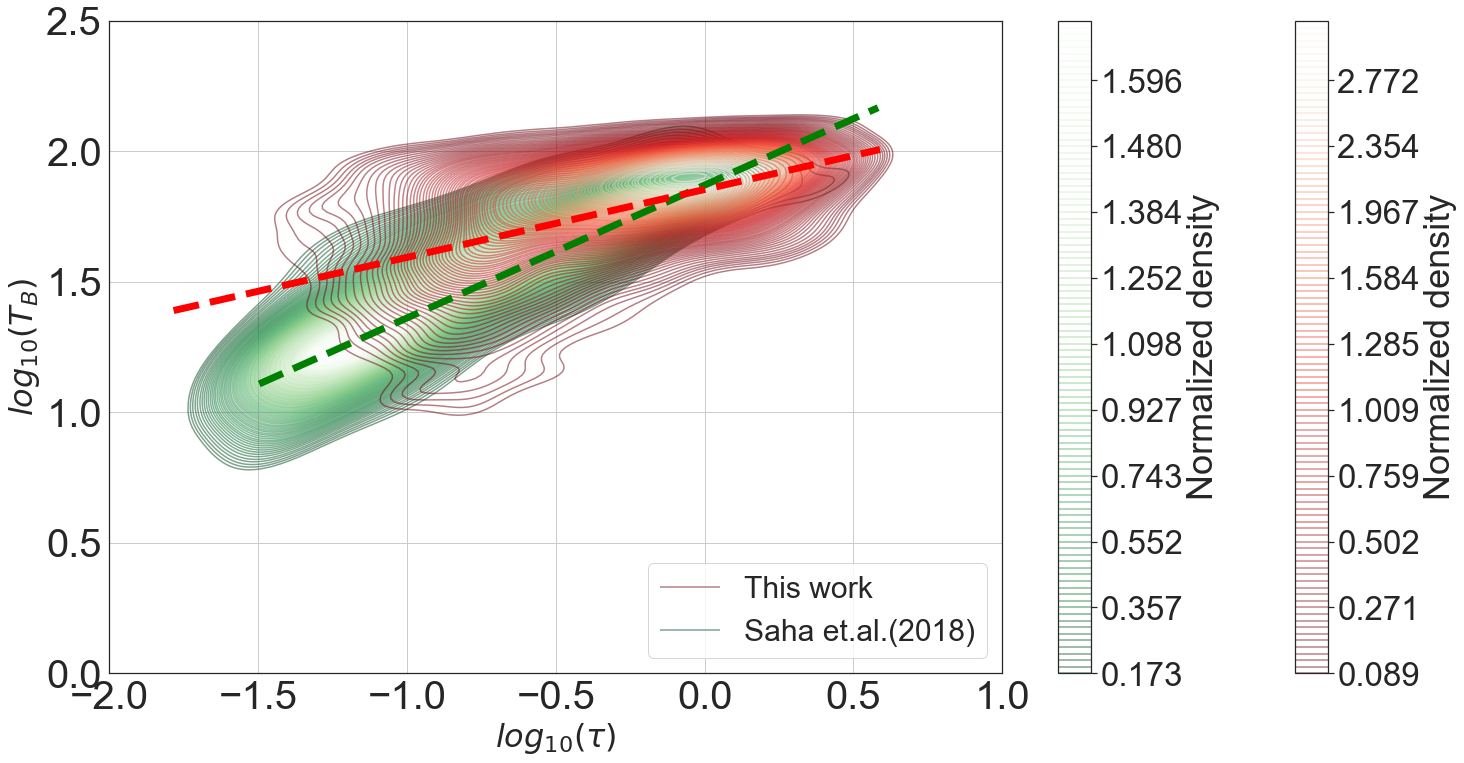}
	\includegraphics[width=60mm]{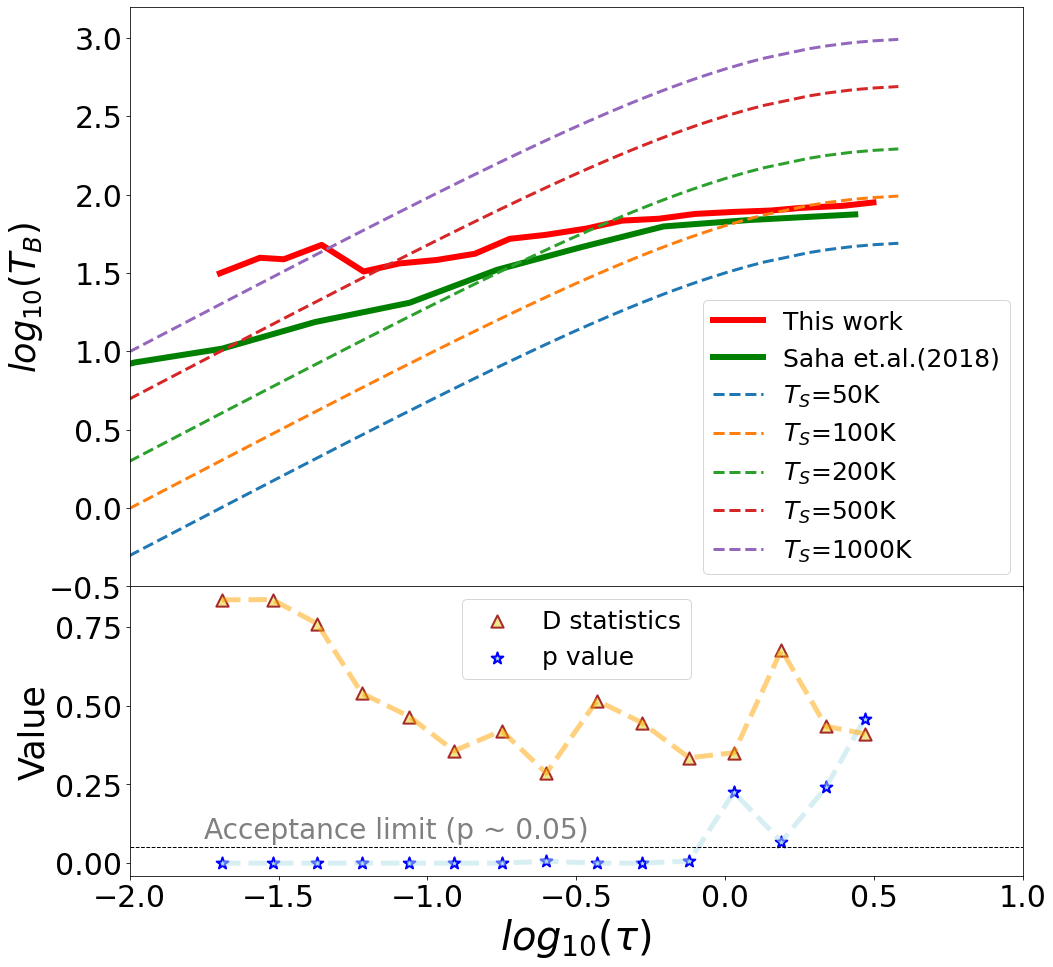}
    \caption{Observed $T_{B} - \tau$ distribution (in $log_{10}$ scale) for H~{\sc i} 21 cm emission and absorption lines. \textbf{Left}: Optical depth ($\tau$) from the THOR and brightness temperature $T_B$ from LAB survey (resampled to the same velocity resolution of 1.5~km~s$^{-1}$) for corresponding spectral channels shown as density contours (red). The comparable $T_{B} - \tau$ distribution for the high latitude sample (ie. \citealt{saha}) is shown in green. Red and green dashed lines are corresponding fitted straight lines for these two samples. \textbf{Right:} (Top panel) Binned $T_B - \tau$ observed median distribution for low (red line) and high latitudes (green line) lines of sight. Dashed lines indicate the $T_B - \tau$ relation for constant $T_{s}$ values. (Bottom panel) Statistical values for the non-parametric K-S test over the same $log_{10}(\tau)$ range.}
    \label{fig:figure1}
\end{figure*}

Galactic H~{\sc i} can have a wide range of temperature, $\sim 20 - 10^{4}$~K \citep{Field,field1969}, with cold gas having higher density and smaller filling factor while the warm gas has lower density and is more diffuse and widespread. Different heating and cooling processes along with self-shielding of the gas yield a temperature of the stable cold neutral medium (CNM) to be $\sim 20 - 400$~K and of the stable warm neutral medium (WNM) to be $\sim 4000 - 10000$~K. Gas with intermediate temperatures is difficult to detect as this phase is expected to be unstable and easily moves to one of the stable phases via runaway heating or cooling (but also see \citealt{heiles2003}, \citealt{audit05}, \citealt{roy2013b}, \citealt{saury14}, and \citealt{murray18} for discussion on gas in the so-called ``unstable'' phase). Combined H~{\sc i} 21 cm emission-absorption observations may be used to study the properties of both the cold and the warm phases \citep{Kulkarni1988,Dickey}. However, inferring properties of the gas in the WNM phase is, in general, more difficult. The optical depth of the WNM is very low due to the lower density and the higher temperature of the gas in this phase. Thus, detecting the shallow and wide components in absorption is challenging and requires a higher sensitivity as well as good spectral baseline stability.

Observational studies have established that for the cold H~{\sc i} ($T_{s} \sim 20 - 300$~K), due to high density ($n \sim 10 - 100$~cm$^{-3}$), the gas is mostly thermalized by collisions (i.e., $T_{s}$ almost equal to $T_{k}$), consistent with theoretical predictions \citep{Clark,radha1972,dickey1978,heiles2003,roy2006}. Additionally, $T_{s}$ and $T_{k}$ may also be coupled radiatively through Lyman-$\alpha$ transition. For the WNM, the relation between $T_{s}$ and $T_{k}$ depends on the details of the physical conditions, but the spin temperature is always expected to be lower than the kinetic temperature \citep{Lizst}.

Another important component of the ISM coexisting with the gas in different phases is dust. Dust takes part in heating and cooling processes of the ISM. It also plays a crucial role in the atomic to molecular transition of the gas as well as in the formation of dense, cold molecular cloud by contributing to the self-shielding. The formation of hydrogen molecules in H~{\sc i} clouds happens efficiently on the surface of the interstellar dust grains due to ionization (and dissociation) of the clouds by near bright sources (mostly stars) \citep{Gould1963,GouldII1963}. Thus it is quite established that the relative amount of gas and dust is indicative of the phase of the gas and is closely related to the formation and evolution of molecular clouds, and eventually of the star formation. 

In this work, we have carried out a systematic study of various properties of the atomic interstellar medium using the H~{\sc i} 21-cm absorption and emission spectra, focusing on a comparison between the gas on the Galactic plane and that at high Galactic latitude. This is enabled by a recent Galactic plane survey \citep{Beuther2016,thor} that provides high quality H~{\sc i} 21 cm absorption spectra towards a large number of background sources probing the lines of sight through the Galactic plane. For a comparison, we have used data from an ongoing absorption survey \citep{roy,roy2013b} probing high Galactic latitude lines of sight. This work concentrates on two aspects - the temperature of the gas and the dust to atomic gas ratio, while comparing the low and the high latitude lines of sight. The paper is organised in the following way: we have provided details of the data used for this work and the analysis procedure in section \S\ref{section:2}. Section \S\ref{section:3} presents the result and a quantitative comparison of the properties inferred from the two samples. We have discussed possible implications of results and presented our conclusions in section \S\ref{section:4}.

\begin{figure*}
    \includegraphics[width=80mm]{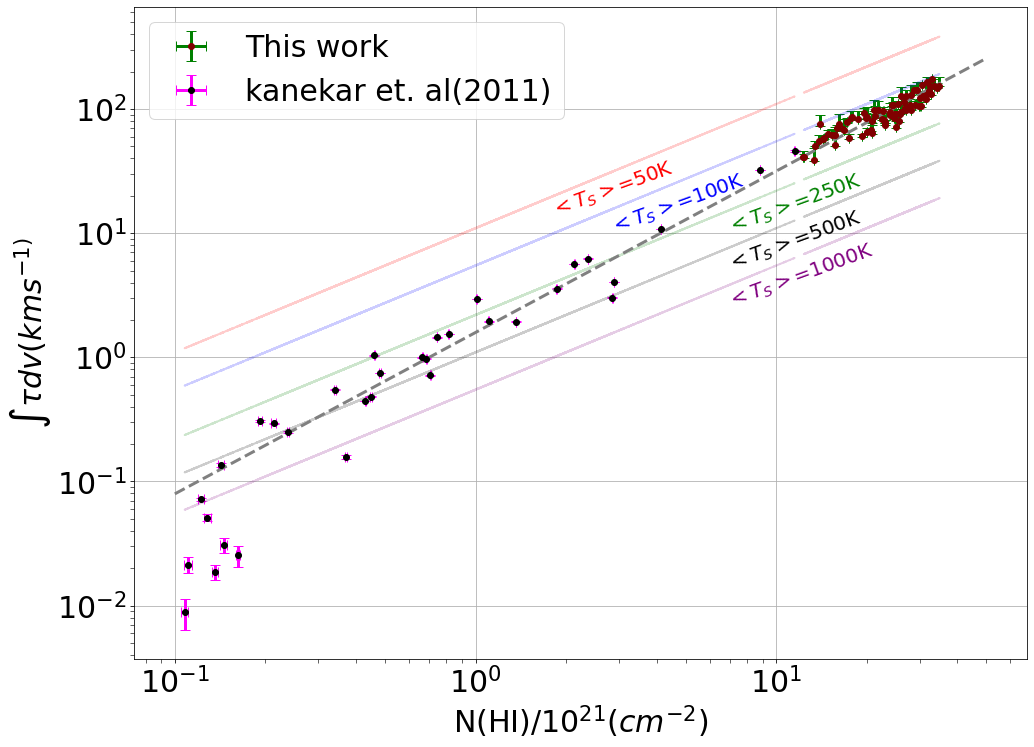}
    \includegraphics[width=84mm]{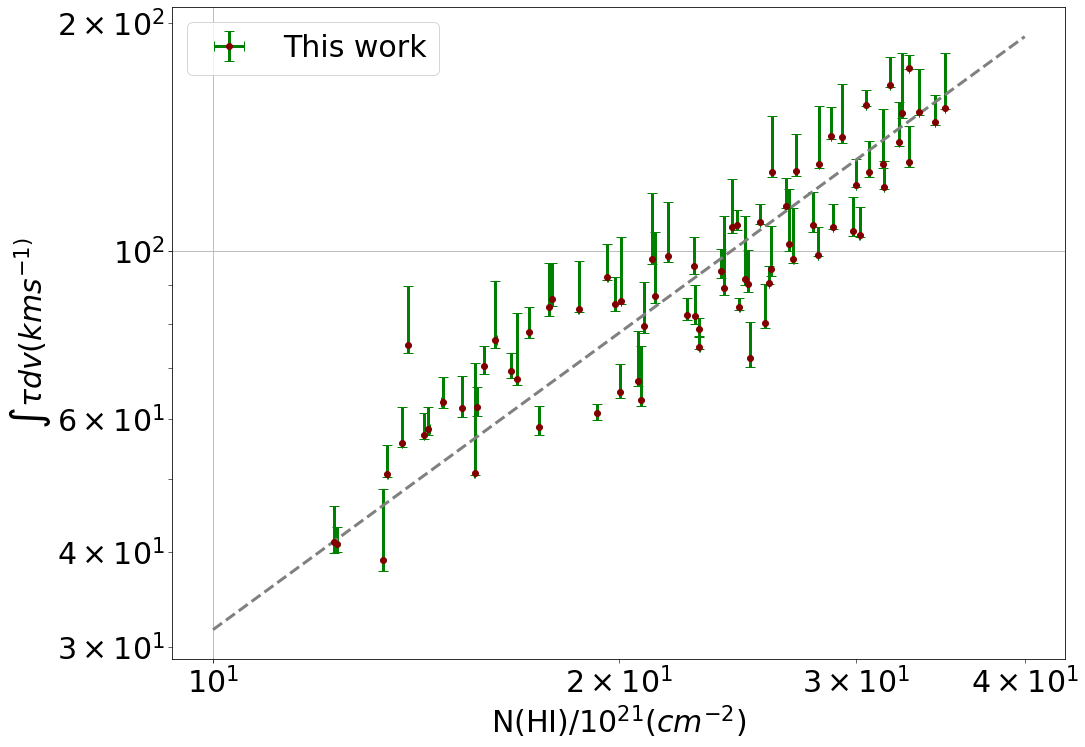}
    
    \caption{Integrated $\tau$ vs $N_{HI}$ column density. \textbf{Left:} The brown points with green errorbars (3$\sigma$) are THOR observations of lines of sight on the Galactic plane (this work), and the black points with  magenta errorbars (3$\sigma$) are lines of sight from H~{\sc i} absorption surveys \citep{roy,kanekar}. The straight lines indicate constant $\langle T_{s} \rangle$ values of $50, 100, 250, 500, 1000$ K (refer to equation \eqref{avgts}). The dotted line is the fitted straight line which pass through all the data for $NHI > 0.15 \times 10^{21} cm ^{-2}$ consistently. \textbf{Right:} Same plot but zoomed into the datapoints (brown points) for the low Galactic latitude sample to show that the linear trend still holds.}
    \label{fig:tauvnhirelations}
\end{figure*}
\section{Data and analysis}
\label{section:2}

With $h{\nu}<<kT$ (where, $h$= Planck's constant, $\nu$= frequency, $k$= Boltzmann's constant, $T$= temperature), using the Rayleigh-Jeans approximation, the specific intensity of H~{\sc i} 21 emission spectra is expressed as HI emission brightness temperature ($T_B = I_{\nu} c^2 /2k {\nu}^2$) as a function of either frequency (${\nu}$) or Doppler velocity ($v$). The 21 cm optical depth $\tau(v)$ can be computed from the absorption spectra $I_{v}$ and the continuum intensity $I_{c}$ of the background source. The quantity ${\tau}={\int} {\alpha} dl $ is the line of sight integral of the linear absorption coefficient ${\alpha}$, which depends on $T_{s}$ and the density of H~{\sc i} ($n_{HI}$). For an isothermal ``cloud'' with known $T_{s}$, the H~{\sc i} column density $N_{HI}$ can be determined from the observed absorption spectrum by integrating $\tau(v)$ over the velocity range of the absorption,
\begin{equation}
    N_{HI} = 1.823\times10^{18} {\int} T_s {\tau} dv \, \, ;
    \label{columndensity}
\end{equation}
where $N_{HI}$ is in cm$^{-2}$, $T_{s}$ is in Kelvin and the velocity interval $dv$ is in km~s$^{-1}$. \citep{Kulkarni1988,Dickey}. The quantities $T_{s}$ and $T_{B}$ are connected through the radiative transfer relation
\begin{equation}
    T_B = T_s [1-exp(-\tau)]\, \, .
    \label{tsrelation}
\end{equation}
Combining equation~\eqref{columndensity} and \eqref{tsrelation}, $N_{HI}$ for an isothermal cloud can be written as
\begin{equation}
    N_{HI} = 1.823\times10^{18} \int \frac{{\tau} T_B}{[1-exp(-\tau)]}\, dv\, \, ;
    \label{NHIrelation}
\end{equation}
in terms of direct observables $T_{B}$ and $\tau$ \citep{dickey1982}. For the optically thin limit ($\tau << 1$), equation~\eqref{NHIrelation} can be simplified to 
\begin{equation}
    N_{HI} = 1.823\times10^{18} \int T_B dv\, \, .
    \label{NHI_optthinrelation}
\end{equation}
Please note, although the observed quantities for the absorption (emission) spectra are $\tau$ ($T_B$) {\it as a function of velocity} in each spectral channel, here the velocity dependence of $T_{B}(v)$, $\tau(v)$ and also $T_{s}(v)$ (from equation~\eqref{tsrelation}) are not shown explicitly. On the other hand, the line of sight average $T_{s}$ can be defined as 
\begin{equation}
    \langle T_s \rangle = \frac{N_{HI}}{1.823\times10^{18} \int \tau \, dv}\, \, ;
    \label{avgts}
\end{equation}
where $N_{HI}$ may be computed from observed spectra using equation~\eqref{NHIrelation}.

\begin{figure*}
    \includegraphics[width=80mm]{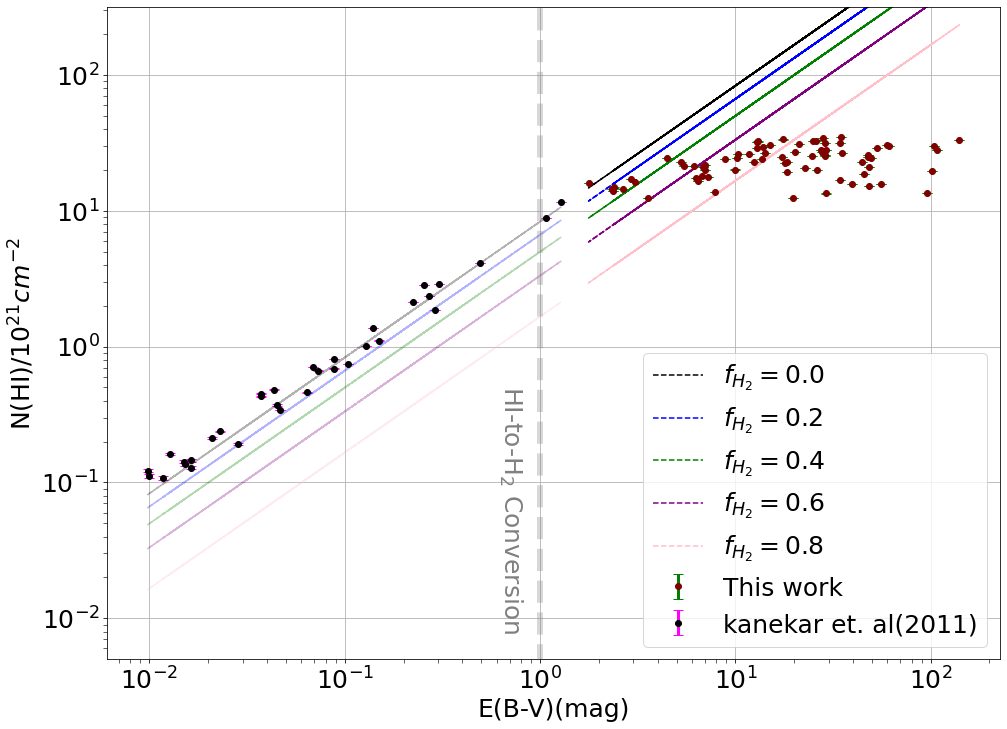}
    \includegraphics[width=82mm]{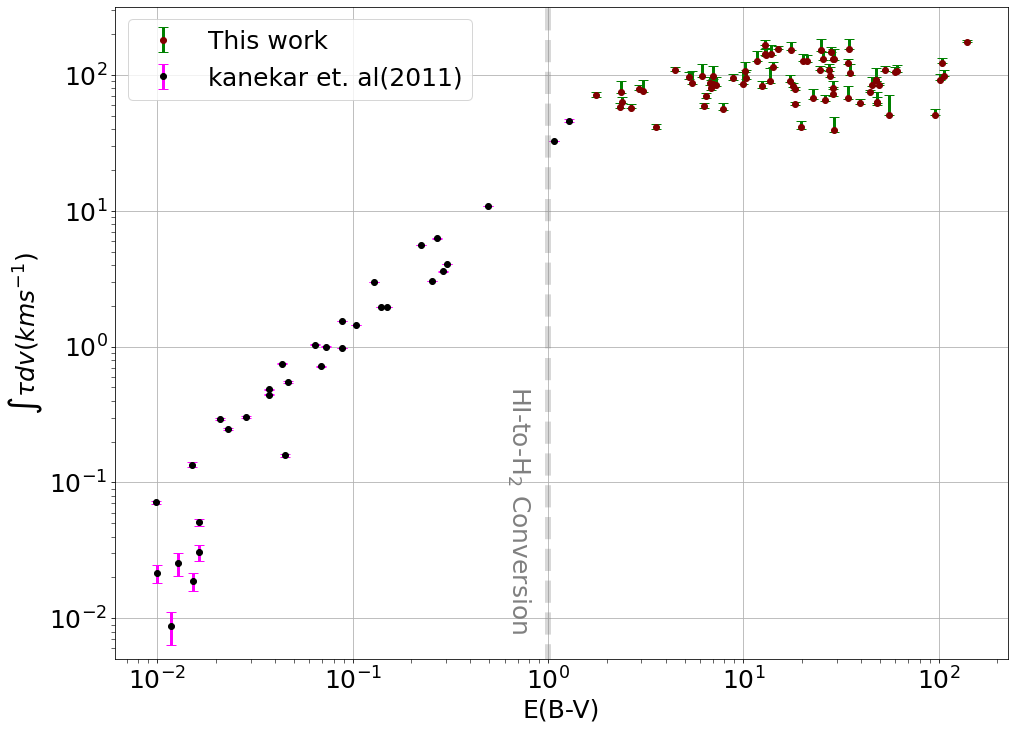}
    \caption{Correlation of $E(B-V)$ with $N_{HI}$ (left panel) and $\int \tau dv$ (right panel). Brown circles with green errorbars (3$\sigma$) are lines of sight on the Galactic plane from THOR (this work), and black circles with magenta errorbars (3$\sigma$) are high latitude lines of sight \citep{roy}. Solid lines (in the left panel) indicate $N_{HI} - E(B-V)$ relation \citep{liszt2014}, $N_{HI} = 8.3\times 10^{21}\,E(B-V)\,(1-f_{H_2})$, for different $H_{2}$ fraction ($f_{H_2}$). Dashed grey line is drawn at E(B-V)=1 where the transition is happening. } 
    \label{fig:extinctionrelations}
\end{figure*}

For studies aiming to understand the properties of the Milky Way ISM, it is relatively easy to have high sensitivity and high spectral resolution H~{\sc i} emission data covering a large area or the full sky from (mostly) single dish or interferometric surveys. But availability of deep H~{\sc i} absorption spectra is limited as suitable bright background sources required for such studies are sparse. One way of estimating, without any bias, the H~{\sc i} column density from only the 21 cm emission spectra has been developed earlier by \citet{saha,chengalur2013}, and has been applied to data for high Galactic latitude lines of sight. 

For this work, we have taken H~{\sc i} absorption spectra from the H~{\sc i}/OH/Recombination line survey of the Milky Way \citep[THOR,][]{Beuther2016,thor}. This is a emission-absorption combined survey that covers part of the Galactic plane ($15\degr < l < 67 \degr$ and $|b| < \pm 1 \degr$), observed with the Karl G. Jansky Very Large Array (VLA) in a C+D+single dish configuration (i.e. an angular resolution of $\approx 40\arcsec$). The emission data is taken from the VLA Galactic Plane Survey (VGPS) \citep{Stil2006} in D and single dish configuration (angular resolution of $\approx 60\arcsec$), whereas, the absorption part is from THOR-only survey done in C-configuration with an angular resolution of $\approx 20\arcsec$ in L-band. The H~{\sc i} data have a spectral resolution of $1.5$ km $s^{-1}$. Along with the H~{\sc i} emission signal, the THOR data also provide H~{\sc i} absorption spectra towards a large number of background sources. We have used these absorption spectra for the current study. For each H~{\sc i} absorption spectra, the optical depth towards the continuum source has been calculated following the prescription described by \citet{bihr2015}. We refer the readers to the original THOR papers for a detailed analysis in deriving the $\tau$ values from the H~{\sc i} absorption spectra.We note that the optical depth RMS for these spectra has a wide range. The spectra with very large optical depth RMS are not of much use as these correspond to the weaker background sources and data from only a few spectral channels will be usable . On the other hand, if we use an RMS cutoff too small, the available number of lines of sight for the analysis will be very small. We have hence used a moderate cutoff and included all spectra with $\tau_{\rm RMS} \leq 0.05$ (i.e., 75 lines of sight) for our analysis. We note that as the $\tau_{\rm RMS}$ is mostly determined by the flux density of the background sources, this cutoff does not introduce any additional bias.

For comparison, we have used absorption spectra from the high spectral resolution and high sensitivity H~{\sc i} absorption survey by \citet{roy}. This survey uses data from the Giant Metrewave Radio Telescope (GMRT) the Westerbork Synthesis Radio Telescope (WSRT), and the Australia Telescope Compact Array (ATCA), with an optical depth RMS sensitivity of $\sim 10^{-3}$ per $1$ km $s^{-1}$ channel, for $32$ lines of sight (excluding the Galactic plane). We have also used complementary 21 cm emission spectra from the Leiden/Argentina/Bonn (LAB) survey \citep[]{LDS,Arnal,Bajaja,Lab} covering $v_{\rm LSR} = -450$ to $+400$ km~s$^{-1}$, at a resolution of $1.3$ km~s$^{-1}$ (angular resolution $\sim 36\arcmin$). We have used spatial interpolation and spectral resampling to get $\tau(v)$ and $T_B(v)$ at the corresponding velocity from the absorption and the emission spectra, respectively.

\begin{figure*}
	\includegraphics[width=54.6mm]{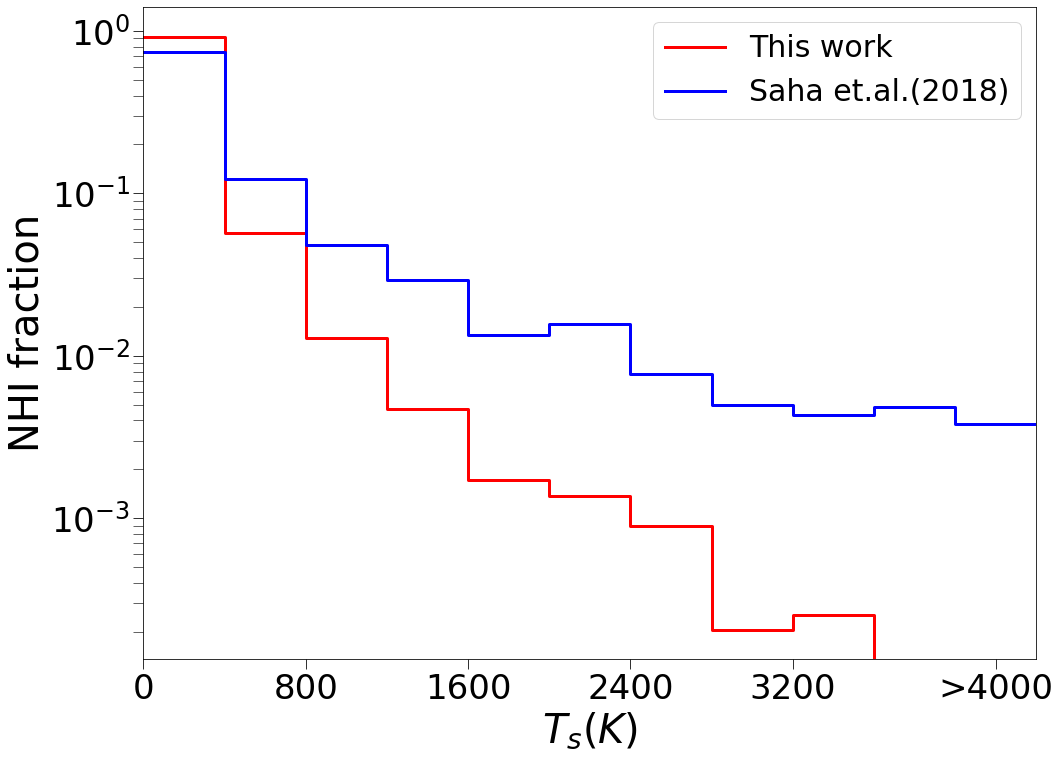} \includegraphics[width=54.6mm]{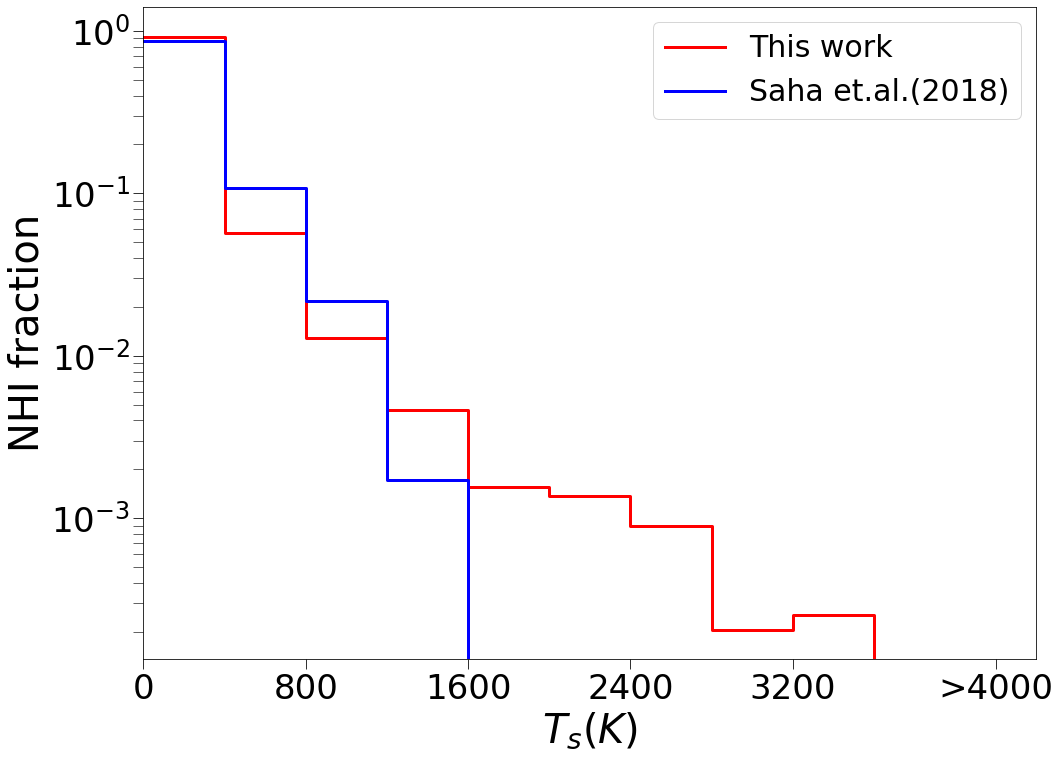}
	\includegraphics[width=54.6mm]{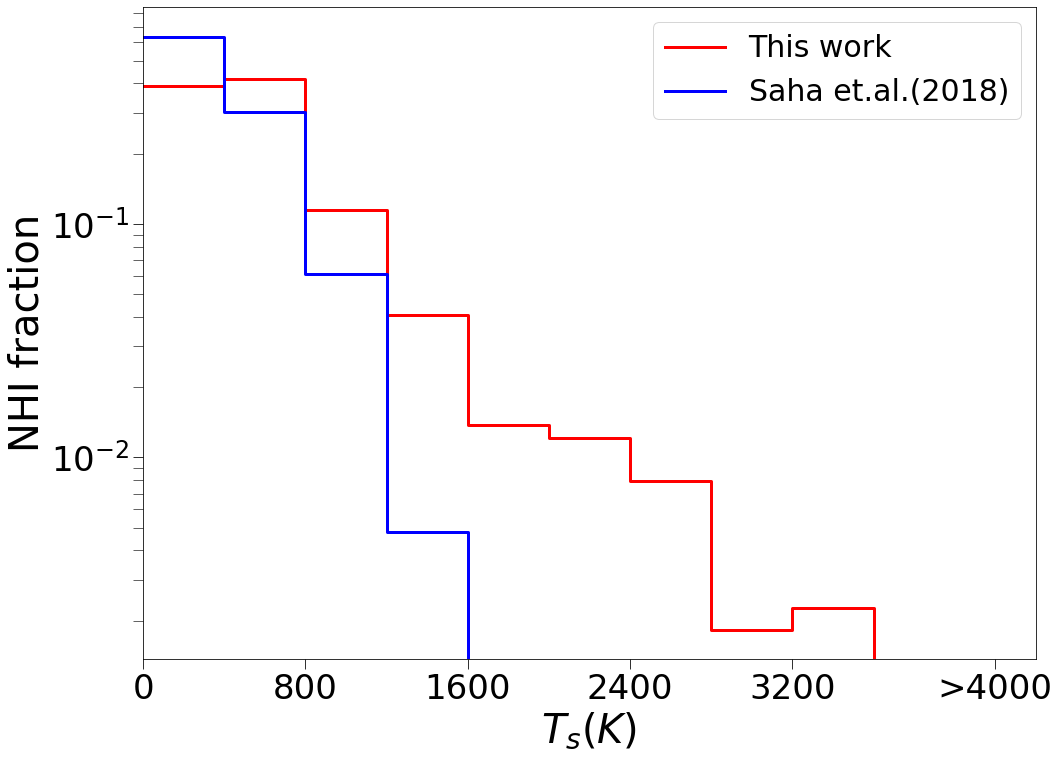}
    \caption{Comparison of H~{\sc i} column density fraction in different $T_{s}$ ranges between the THOR sample (this work) and the high latitude sample \citep{roy,saha}, shown in red and blue line respectively. \textbf{Left panel}: Entire range of $\tau$; \textbf{Middle panel}: For $\tau > 0.02$ \textbf{Right panel}: For $0.2 > \tau > 0.02$. }
    \label{fig:tsdistribution}
\end{figure*}

\section{Results}
\label{section:3}

For the subsample of 75 lines of sight with reasonable S/N from THOR, we show the distribution of $T_B(v)$ and $\tau(v)$ from individual channels in Figure~\ref{fig:figure1}. The plotted density contours in the left panel are from observed data points above $3\sigma$, with the mean and the median values for different $\tau$ bins are shown in dashed and solid red lines, respectively. The same for the high latitude sample is shown in green lines for a comparison. Due to the higher sensitivity (lower $\tau{\rm RMS}$) of the latter sample, the distribution (truncated in the figure, but see \citealt{saha}) is extended to much lower $\tau$ (and $T_B$), whereas a large number of data points for the Galactic plane sample is at higher optical depth. We note here that there is a small fraction of data points from THOR with optical depth saturation, where we only have a lower limit to the optical depth. This however, does not change the $T_B - \tau$ relation shown in Figure~\ref{fig:figure1}, apart from extending the flat part to even higher optical depth. This concentration of data points at the high $\tau$ end of the distribution is consistent with a large fraction of cold, dense gas in the Galactic plane. This is evident from the top right panel of Figure~\ref{fig:figure1}, which shows the median $T_B - \tau$ for both samples along with the constant $T_{s}$ lines (from equation~\eqref{tsrelation}), with the high $\tau$ end having lower $T_{s}$. However, at the intermediate $\tau$ range ($0.02 \lesssim \tau \lesssim 0.2$), we found the two distributions to deviate, showing relatively higher $T_B$ (i.e, higher $T_{s}$ as well) for the lines of sight on the Galactic plane from THOR. To establish this deviation, we have performed a least square fitting (shown in Figure~\ref{fig:figure1} left panel)  in both the samples considering a linear function in the log-log scale and the fitted models (corresponding parameters for high latitude sample: slope = $0.509 \pm 0.020$ and intercept = $1.871 \pm 0.019$ ; for THOR sample: slope = $0.260 \pm 0.007$ and intercept = $1.854 \pm 0.004$ ) do not match within their respective errors. The trend remains the same when we carried out the similar parametric least square fitting with an extra quadratic term in the function (ie. the fitted slope terms are well beyond their individual uncertainties). In addition to that, we have also performed the non-parametric Kolmogorov-Smirnov (K-S) test for different bins in optical depth (in log scale) and shown the results (D-statistics and p value) in the bottom right panel in Figure~\ref{fig:figure1}. Clearly, below $\tau<1.0$ the two samples are very unlikely to be drawn from the same underlying distribution. Both of these parametric and non-parametric tests strongly support the inference of a statistically significant difference between the two samples. There are almost $15\%$ (precisely $14.89\%$) channels that show optical depth saturation, but they are above $\tau \sim 1.3$ , and should not affect the K-S statistics or the slope of the $T_{B} - \tau$ relation below $\tau<1.0$.

It is interesting to note that even if the channels with intermediate $\tau$ have a higher $T_{s}$ for the THOR sample, the line of sight average spin temperature ($\langle T_{s} \rangle$) is lower than that of the high latitude lines of sight \citep{saha,roy}. In fact, as shown in Figure~\ref{fig:tauvnhirelations}, the THOR data points on the $N_{HI} - \int\tau\,dv$ plot follow the same trend of lower $\langle T_{s} \rangle$ for higher $N_{HI}$ \citep{kanekar,roy}, even in the zoomed scale.  Note that NHI from LAB emission spectra is computed for the matching velocity range over which THOR absorption spectra are also integrated. The errors for the THOR sample have been calculated with a simple prescription considering the optical thin approximation in a $3 \sigma$ limit and the errors for the high latitude sample are taken from \cite{kanekar} (see Table 1 in that paper). The contributions for the saturation points are incorporated by including an extra uncertainty, corresponding to a typical factor of two higher optical depth for those channels, in the upper limits of the errorbars (which are more readily visible in the zoomed plot in the right panel). For $N_{HI} > 10^{22}$~cm$^{-2}$ we find $\langle T_{s} \rangle \lesssim 100$~K, whereas for $N_{HI} <  2\times 10^{20}$~cm$^{-2}$, as reported in \citet{kanekar}, $\langle T_{s} \rangle \gtrsim 600$~K.

Apart from the lower average temperature for the atomic gas, we also expect the molecular gas fraction to be significantly higher in the Galactic plane. One way to explore this is to check the correlation between the H~{\sc i} column density and the colour excess $E(B-V)$. Earlier observations have revealed a tight linear relation between hydrogen column density and $E(B-V)$ \citep{bohlin1978,savage1977,liszt2010,liszt2013,liszt2014,heiles1976,burstein1982,schegel,Lab}. A tight correlation between $E(B-V)$ and the integrated optical depth is also reported by \citet{liszt2014}. For the lines of sight in both the samples, we have taken the $E(B-V)$ values(angular resolution $\sim 6\arcmin$) derived from the recalibrated infrared dust maps, from the NASA/IPAC Extinction calculator \citep{schegel,sf11}. In Figure~\ref{fig:extinctionrelations}, these correlations are shown for both the THOR sample and the high Galactic latitude sample. The straight lines in the left panel are $N_{HI} - E(B-V)$ relation from different molecular gas fractions ($f_{H_2}$) taken from \citet{liszt2014}. In both cases, we see clear flattening with a sharp transition of the relations at about $E(B-V) = 1$ (marked with dashed grey line), incidentally coinciding with the separating $E(B-V)$ for the two samples. The observed high $E(B-V)$ is expected due to the abundance of large amount of dust on the Galactic plane. Even if the H~{\sc i} column density can be somewhat underestimated due to the high optical depth and absorption saturation of some spectral channels, the sharp break and the flattening must be primarily due to the atomic-to-molecular transition of the ISM and a higher molecular gas fraction on the Galactic plane at $E(B-V) > 1$. Unlike \citet{liszt2014} \& \citet{kanekar}, we see the flattening at $\int \tau dv > 100$. The plausible reason behind this trend is that the column density here is already corrected for the optical depth using the isothermal estimator of $N_{HI}$. Note that, we see this flattening in $N_{HI} \sim 10^{22} cm^{-2}$, which is consistent with the understanding that the molecular fraction is significant for $N_{HI} > 10^{22} cm^{-2}$ \citep{schaye2001,Krumholz2009}. 

To reconcile these two apparently contradictory results, viz. lower $\langle T_{s} \rangle$ but higher $T_{s}(v)$ for intermediate $\tau$ range data points coming from the low Galactic latitude lines of sight of THOR, we have looked at the distribution of $T_{s}(v)$ for the two samples more closely. This is done by computing the $T_{s}$ and the H~{\sc i} column density for each individual channel, and determining the $N_{HI}$ fraction for different $T_{s}$ bins. As expected, for both samples, most of the gas has low $T_{s}$ ($< 400$~K). For the THOR sample, the fractions of $N_{HI}$ for $T_{s}<200$~K , $200$~K $<T_{s}<1000$~K and $T_{s} > 1000$~K are $\sim 74\%$, $25\%$ and $1\%$, respectively. For the high latitude sample, \citet{roy} reported the corresponding fractions to be $46\%$, $37\%$ and $9\%$. As shown in the left panel of  Figure~\ref{fig:tsdistribution}, the high latitude sample indeed shows a larger fraction of warm gas compared to that for the Galactic plane data, increasing $\langle T_{s} \rangle$ for those lines of sight. This is, however, considering all the data points including the very low optical depth channels for the high latitude sample. Instead, if we use an optical depth cutoff of $\tau > 0.02$ (Figure~\ref{fig:tsdistribution}, middle panel) or $0.2 > \tau > 0.02$ (right panel), we see a clear excess of warmer gas in the low Galactic latitude data from THOR compared to the high latitude data from the earlier absorption survey. The similar trend in the spin temperature is also in the other works by \citet{heiles2003,nguyen2019,Murray2021}.

\section{Discussion and Conclusions}
\label{section:4}

\begin{figure}
\includegraphics[width=\columnwidth]{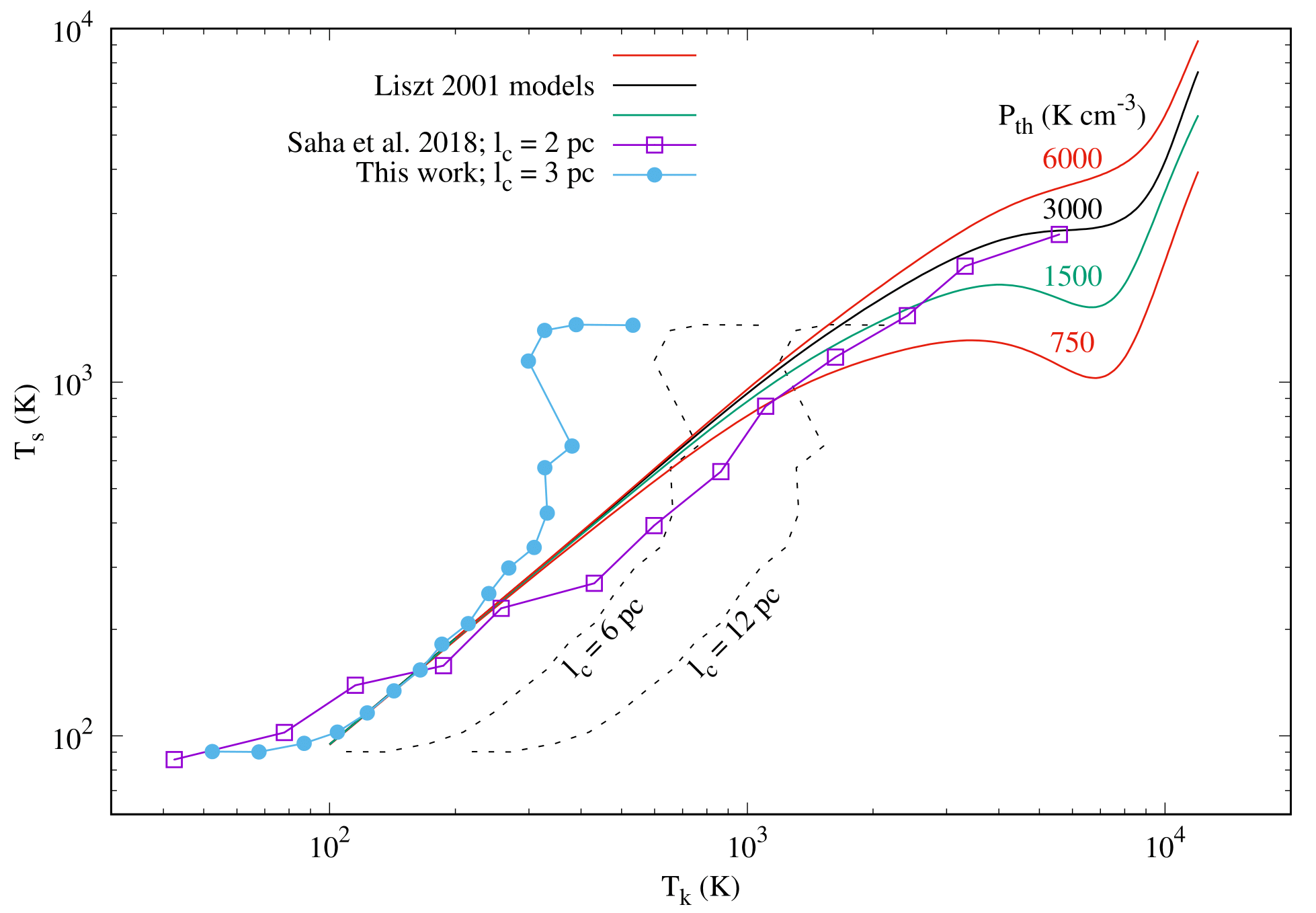}
\caption{Comparison of inferred $T_{k}$ and $T_{s}$ from the THOR data with models from \citet{Lizst}. Solid lines are model $T_{k} - T_{s}$ relation for different thermal pressure. Line with filled circles is from this analysis, for $P_{th} = 3000 \,K cm^{-3}$ and $l_c = 3 \,pc$. Dashed lines are for $l_c = 6$ and 12 pc. Line with empty squares is based on data from \citet{saha} for $l_c = 2\, pc$ for a comparison with the low latitude sample.}
    \label{fig:tstkcompare}
\end{figure}

We want to emphasize the important point here that the $T_{s}$ distribution shown here comes from measured $T_B$ and $\tau$ in individual channels; the inferred $T_{s}$ for a channel is an average of multiple components with different individual $T_{s}$ blended at that particular Doppler velocity, and may not be spatially coexisting. Hence, an observed intermediate value of $T_{s}$ does not necessarily imply the presence of gas in the thermally unstable phase. However, this is true for both the samples, and the observed higher $T_{s}(v)$ at intermediate $\tau$ for the THOR sample indicates a real difference in physical properties for the gas on the Galactic plane compared to that along high latitude lines of sight.

For a single spectral channel, $T_{s}(v) = T_{B}(v)/\tau(v)$ can be expressed as $[(f_C/T_{s,C})+(f_W/T_{s,W})]^{-1}$ in terms of the column density fraction of cold ($f_C$) and warm gas ($f_W = 1 - f_C$) for that channel, and the spin temperature of cold and warm gas, $T_{s,C}$ and $T_{s,W}$, respectively. For a given $\tau$, higher $T_B$ (i.e. higher $T_{s}$) can be due to higher $T_{s,C}$, $T_{s,W}$ and/or lower $f_C$. However, as the average $T_s$ for the channel is heavily biased to that of the cold phase [ For eg. even for a $50\%-50\%$ mixture of cold gas( $\sim 100 K$) and warm ( $\sim 5000K$), average $T_s$ remains in the colder phase ( < 200 K) ], explaining higher observed $T_s$ requires either significantly higher $T_{s,W}$ or very small $f_C$ - both of which seem unlikely for the Galactic plane gas. Alternatively, somewhat higher $T_{s,C}$ for a small fraction of gas can easily result in a higher average $T_s$ for a velocity channel, and that seems to be a more plausible scenario. 

To understand the small fraction of high $T_{s}$ gas with relatively high optical depth, we further convert the observed $T_B - \tau$ relation to $T_{k} - T_{s}$ relation under some simplifying assumption, and compare it with theoretical expectations. From the binned $T_B$ and $\tau$, we get the corresponding $T_{s}$ for the bin using equation (2). We can also compute $T_{k}$ assuming a thermal pressure $P_{th}$ and characteristic length scale $l_c$ (corresponding to the spectral resolution) as
\begin{equation}
    T_k = P_{th}/n_{HI} = P_{th}l_c/N_{HI}
    \label{ennpl}
\end{equation}
where $N_{HI}$ is given by equation (3). Here, $P_{th}$ and $l_c$ (or, more appropriately $P_{th}l_c$) are parameters that may be adjusted to see if the observed $T_{k} - T_{s}$ relation can be matched with the theoretical one \citep[e.g.][]{Lizst}. As shown in Fig ~\ref{fig:tstkcompare}, for a nominal choice of $P_{th} = 3000 \,K cm^{-3}$, if we set $l_c = 2 \,pc$ (per $1$~ km~s$^{-1}$ velocity width), the high latitude data match well with the theoretical curve, except for very low $T_{s}$ end. The deviation of the low $T_{s}$ end can be explained either by a higher pressure or a larger value of $l_c$. However, for the Galactic plane sample, we need a slightly higher $l_c = 3 \,pc$ to match the data with expected models for $100$~K $< T_{s} < 300 $~K, and significantly higher values (up to 12 pc per $1$~ km~s$^{-1}$ velocity width) for $T_{s} > 300$~K bins. As there is a degeneracy between $P_{th}$ and $l_c$ in this analysis, part of the discrepancy may be explained as originating from higher thermal pressure in the Galactic plane compared to the high latitude sight lines; however, particularly for the high $T_{s}$ end, there is clear indication of a larger length scale (equivalently, a lower number density) to account for the deviation. 

In summary, we have carried out a systematic comparison of the THOR H~{\sc i} absorption spectra with the comparable data for a high Galactic latitude sample. As expected, the gas in the Galactic plane has lower average spin temperature, higher dust extinction and higher molecular gas fraction. There is an excess, compared to the high latitude lines of sight, of gas at relatively higher spin temperature at intermediate optical depth range. This is likely to be a small fraction of cold gas at slightly higher temperature and lower density on the Galactic plane. It will be interesting to further explore the origin and properties of this high $T_{s}$ gas with future high sensitive measurements of optical depth for low latitude lines of sight passing through the Galactic plane.


\section*{Acknowledgements}

We thank the scientific editor and the anonymous referee for their valuable comments which have helped us improving the presentation of the work. This research has made use of NASA’s Astrophysics Data System and the NASA/IPAC Extragalactic Database. Data used in this paper were obtained from the Leiden/Argentina/Bonn Galactic H~{\sc i} survey, THOR survey, and the ATCA/GMRT/WSRT H~{\sc i} absorption survey. The National Radio Astronomy Observatory is a facility of the National Science Foundation operated under cooperative agreement by Associated Universities, Inc. HB acknowledges support from the European Research Council under the Horizon 2020 Framework Programme via the ERC Consolidator Grant CSF-648505. HB and JS also acknowledge support from the Deutsche Forschungsgemeinschaft in the Collaborative Research Center (SFB 881 - Project-ID 138713538) "The Milky Way System" (subproject B1).  We acknowledge Dr. Yuan Wang, Dr. Kanan K. Datta for their useful comments and providing the opportunity to initiate this project. 

\section*{Data Availability}
The data used for this study is available from the Leiden/Argentine/Bonn Galactic H~{\sc i} survey, THOR survey, and the ATCA/GMRT/WSRT H~{\sc i} absorption survey. The final data products from this study will be shared on reasonable request to the corresponding author.



\bibliographystyle{mnras}
\bibliography{mnras}




%
%


\bsp	
\label{lastpage}
\end{document}